\documentclass[DIV=calc,paper=a4,fontsize=10pt,twocolumn,abstracton]{scrartcl}
\usepackage[font=small,format=plain,labelfont=bf,up,textfont=it,up]{caption}
\usepackage[numbers]{natbib}
\usepackage{abstract}
\usepackage{aas_macros}
\usepackage{epsfig}
\usepackage{graphicx}

\DeclareRobustCommand{\ion}[2]{%
\relax\ifmmode
\ifx\testbx\f@series
{\mathbf{#1\,\mathsc{#2}}}\else
{\mathrm{#1\,\mathsc{#2}}}\fi
\else\textup{#1\,{\mdseries\textsc{#2}}}%
\fi}

\def\druck@lement#1{{}^{\2}_{\3}\mathrm{#1}{}^{\1}_{\4}{}\if@tempswa$\fi$}

\begin{document}

\title{AGN Obscuration and the Unified Model}
\author{Stefano Bianchi\thanks{Dipartimento di Fisica, Universit\`a degli Studi Roma Tre, via della Vasca Navale 84, 00146 Roma, Italy}, Roberto Maiolino\thanks{INAF-Osservatorio Astronomico di Roma, via di Frascati 33, 00040 Monte Porzio Catone, Italy} \thanks{Cavendish Laboratory, University of Cambridge, 19 J. J. Thomson Ave., Cambridge CB3 0HE, UK}, Guido Risaliti\thanks{INAF-Osservatorio Astrofisico di Arcetri, L.go E. Fermi 5, Firenze, Italy; and Harvard-Smithsonian Center for Astrophysics, 60 Garden Street, Cambridge, MA 0218, USA}}
\date{}

\twocolumn[
\maketitle 
\begin{onecolabstract}
Unification Models of Active Galactic Nuclei postulate that all the observed differences between Type 1
and Type 2 objects are due to orientation effects with respect to the line-of-sight to the observer. The
key ingredient of these models is the obscuring medium, historically envisaged as a toroidal structure on
a parsec scale. However, many results obtained in the last few years are clearly showing the need for a
more complex geometrical distribution of the absorbing media. In this paper we review the various pieces of
evidence for obscuring media on different scales, from the vicinity of
the black hole to the host galaxy, in order to picture an updated unification scenario explaining the complex observed phenomenology. We conclude by mentioning some of the open issues.
\vskip 1cm
\end{onecolabstract}
]
\saythanks      

\section{\label{introduction}Introduction: the standard Unified Model}

In this review, we discuss the recent developments on the AGN
unified models, specifically for what concern the geometry, location
and physics of the absorbing medium. Before discussing the more
recent results, in this section we shortly review the
early, classical arguments that historically
led to the formulation of the
standard Unified Model. An early review of the initial results was also given
in \citet{antonucci93}.

The first unification attempts have been focussed on polarization measurements. In particular \citet{antonucci84} found a perpendicular alignment of optical polarization relative to the radio axis in a sample of radio galaxies, which was interpreted as due to scattering of photons, whose direction before entering the line of sight was primarily in the vertical direction.
Shortly after, additional evidence was found in low luminosity, local
AGN, and specifically Seyfert galaxies \citep{sey43}. Seyfert 1 galaxies are characterized
by the presence of broad optical
permitted lines (FWHM$>$1000 km/s), such as H$\alpha$
and H$\beta$, that are not observed in Seyfert 2 galaxies. However, the
presence of both strong high ionization and low ionization
narrow (FWHM$<$1000 km/s) forbidden lines (such as [\ion{O}{iii}], [\ion{Ne}{iii}],
[\ion{O}{ii}], [\ion{O}{i}], [\ion{N}{ii}], [\ion{S}{ii}]), and several very high ionization
coronal lines (such as [\ion{Fe}{x}], [\ion{Fe}{xi}], [\ion{Si}{ix}], [\ion{Si}{x}])
is common to both types of Seyfert
galaxies and with similar line ratios. The latter finding suggested that
all Seyfert galaxies are powered by the same intrinsic engine.

A strong observational evidence of a unification between type 2 and
type 1 Seyfert nuclei has been the
discovery of broad optical lines in the polarized spectrum of the
archetypal Seyfert 2, NGC~1068, obtained by \citet{antonucci85}. This finding
revealed the presence of a Broad Line Region (BLR) in this Seyfert 2 nucleus,
which is hidden to our line of sight, but whose light is scattered in our
direction from material (in the case of NGC~1068, likely free electrons in
ionized gas) distributed on scales larger than the absorber. Such reflected
light is very weak compared to the light of the galaxy, hence difficult to
detect in the total spectrum, but it is highly polarized, and therefore detectable in
the polarized spectrum. The basic idea of the Unified Model, is that
type 2 and type 1 AGN are intrinsically the same class of objects and their
differences are only due to orientation effects relative to an obscuring
medium.

The need for a ``toroidal'', axisymmetric structure of the absorber
was initially inferred 
from the fact itself that the reflected broad lines are polarized and from the measured polarization angles.
If the absorber was a simple cloud along the line of sight, then reflection
should come from all directions, hence by averaging from all angles the total
polarization should be zero. In order to break the symmetry of the polarization
angles the absorber should prevent the nuclear light to be scattered in a
significant range of angles, and a ``torus'' is the most natural configuration
that can achieve this effect.

The size of the toroidal absorber was initially postulated to be on the parsec
scale \citep{kb86,kb88}. Such typical size was simply inferred by the need for the absorber
to be large enough to obscure the BLR, which in Seyfert nuclei
has a size well below a parsec, based on reverberation studies \citep[e.g.][]{kaspi05,peterson93}, but small enough not to obscure the Narrow Line
Region (NLR), which is distributed on the 10-100 pc scale. However, as
we will discuss extensively in this paper, there is clear evidence that the
absorbing medium is also distributed on smaller and larger scales.

Since the seminal discovery of \citet{antonucci85}, polarized broad lines
have been discovered in several other Seyfert 2 nuclei
\citep{tran01,tran03}, contributing to the generalization
of the unified model to the whole class of type 2 AGN. 
Further evidence for a unified theory between type 2 and type 1 AGN has
been obtained from various multiwavelength studies.
In a number of type 2 AGN the presence of an obscured BLR was inferred
from the detection of a broad component of hydrogen recombination lines in the near
infrared, such as Pa$\beta$, Pa$\alpha$, Br$\gamma$, where dust absorption
can be several times lower than in the optical \citep{veilleux97,nagar02,
lutz02,riffel06,cai10}.

Hard X-ray observations of AGN have provided additional, unambiguous
evidence in favor of the unified model, obtained already with the early
hard X-ray satellites such as Exosat, Ginga, ASCA, BeppoSAX 
\citep{sd96,turner97,mai98,bassani99}.
Although generally weak or even undetected in the soft X-ray band ($<$2~keV),
most type 2 AGN are detected in the hard X-rays ($>$2~keV) and are
characterized by a power-law spectrum similar to Sy1s, favoring a common
central engine, but affected by a photoelectric absorption cutoff directly
demonstrating the presence of an absorbing medium along the line of sight,
with a column density typically well in excess of $\rm 10^{22}~cm^{-2}$.
In a number of Seyfert 2s the hard X-ray spectrum does not show the presence
of a prominent absorption cutoff, but it is characterized by a very strong
Fe K$\alpha$ line at 6.4~keV, with an equivalent width larger than 500~eV
\citep[e.g.][]{matt97,matt99,bassani99}.
Such high equivalent widths of the Fe line can only be explained by assuming
that the direct X-ray radiation is totally absorbed by a Compton thick
medium ($\rm N_H > 10^{24}~cm^{-2}$) and that the observed (weak) continuum
and Fe K$\alpha$ are due to reflection from the circumnuclear medium.
A detailed discussion of these effects and on the distribution of the
reflecting medium will be given in the next sections.

If the NLR in type 2 AGN is created by the photoionization
of a nuclear UV/X source hidden from our line of sight by a toroidal
absorber, then the expectation is that the NLR should have a (bi-)conical morphology,
due to the light cones defined by the nuclear absorber.
High resolution, narrow band imaging (or integral field spectroscopy),
especially with the advent of HST, have indeed revealed such ionization
cones on scales ranging from a few 10~pc up to several 100~pc,
in many nearby AGN \citep[e.g.][]{pogge88,evans91,wt94,
barbosa09}.
The opening angle of the cones gives the fraction of the sky hidden to
our line of sight, which is in reasonable agreement with what inferred
from the relative fraction between type 1 and type 2 AGN in the
local universe \citep{mr95}. It is interesting that the axis of the
ionization cones is often not aligned with the minor axis of the host galaxy,
meaning that the circumnuclear absorber is not necessarily aligned with the
gaseous disk of the host galaxy. Generally the orientation of the
ionization cone axis is in the same direction as
the radio jet, but often the two are not exactly aligned, meaning
that even on small (parsec or sub-parsec) scales there is a slight
misalignment between the dusty absorbing medium and the central engine
(the accretion disc) \citep{wt94,nag99}.

We conclude this introduction by shortly mentioning that several theoretical works
have modelled the physics and the structure of the ``toroidal''
absorbing medium in the attempt of reproducing the observable properties.
Initial models had assumed a simple toroidal structure with a uniform
distribution of gas and dust with a parsec-scale radius
\citep[e.g.][]{kb88,pier92}, while other models suggested more extended
geometries, up to 100 pc, to explain the broad infrared spectral energy
distribution observed in AGN \citep{granato97}.
One of the main problems of these models
is their dynamical stability. Often radiation pressure from infrared photons
within the torus is invoked as a solution to keep the torus geometrically
thick \citep{krolik07}. Other authors ascribe the geometrical thickness
of the torus to turbulence introduced by supernovae or stellar winds
\citep{wada05,watabe05}. It has also been proposed that
a wide angle of obscuration does not necessarily require a geometrically
thick torus, but can also be achieved with a warped or tilted disk
\citep{nay05,caproni06a,caproni06b,lawrence10}.

Recently, the hypothesis of uniform gas and dust distribution has been
abandoned by many models, by introducing a clumpy structure of the absorbing
medium \citep{es06,nenkova08,honig07,honig10}. These models can
account for several of the observational properties of AGN and, most
importantly, are strongly supported by recent X-ray
observations that directly reveal the clumpiness of the absorbing medium,
as discussed in the following sections.

\section{From Galactic to sub-pc scale: absorption at different scales}

One of the most significant new aspects on the structure of AGN, as emerged in last few years, is that
the standard, parsec-scale ``torus'' is not enough to explain all the complex absorption features
discovered by many observations. While the unified picture remains valid in its more general sense (i.e.
the presence of non-spherically symmetric absorbers at the origin of the type 1/type 2 dichotomy) several
new observations and models, mostly in the X-ray and infrared domain, suggest that multiple absorbers are
present around the central source, on quite different physical scales. In the following we review the main observational evidence for each of them, together with some brief discussion on their physical interpretation.

\subsection{\label{subdust}Absorption within the sublimation radius}

The evidence for gas absorption within the sublimation radius comes mostly from X-ray observations. The most direct way to probe the presence of such gas component is through absorption variability measurements.

X-ray absorption variability is a common feature in AGN. An analysis of a sample of nearby obscured AGN with multiple X-ray observations, performed a few years ago \citep{risa02} revealed that column density (N$_H$) variations are almost ubiquitous in local Seyfert galaxies. More recent observations performed with XMM-\textit{Newton}, {\em Chandra} and {\em Suzaku} further confirmed this finding. The physical implications of these measurements  are that the  circumnuclear X-ray absorber (or, at least, one component of it) must be clumpy, and located at sub-parsec distances from the central source. 

The comparison between different observations, typically performed at time distances of months-years,
only provides upper limits to the intrinsic time scales of $N_H$ variations. An improvement of  these
estimates could only be obtained through observational campaigns within a few weeks/days, and/or through the search for N$_H$ variations within single long observations. Such short time-scale studies have been 
performed for a handful sources: NGC~1365 \citep{ris05,ris07,ris09a,mai10}, NGC~4388 \citep{elvis04}, NGC~4151 \citep{puc07}, NGC~7582 \citep{bianchi09c}, Mrk~766 \citep{risa11}. 

\begin{figure*}
\centering
\hbox to \textwidth
{
\parbox{0.45\textwidth}{
\includegraphics[width=0.45\textwidth]{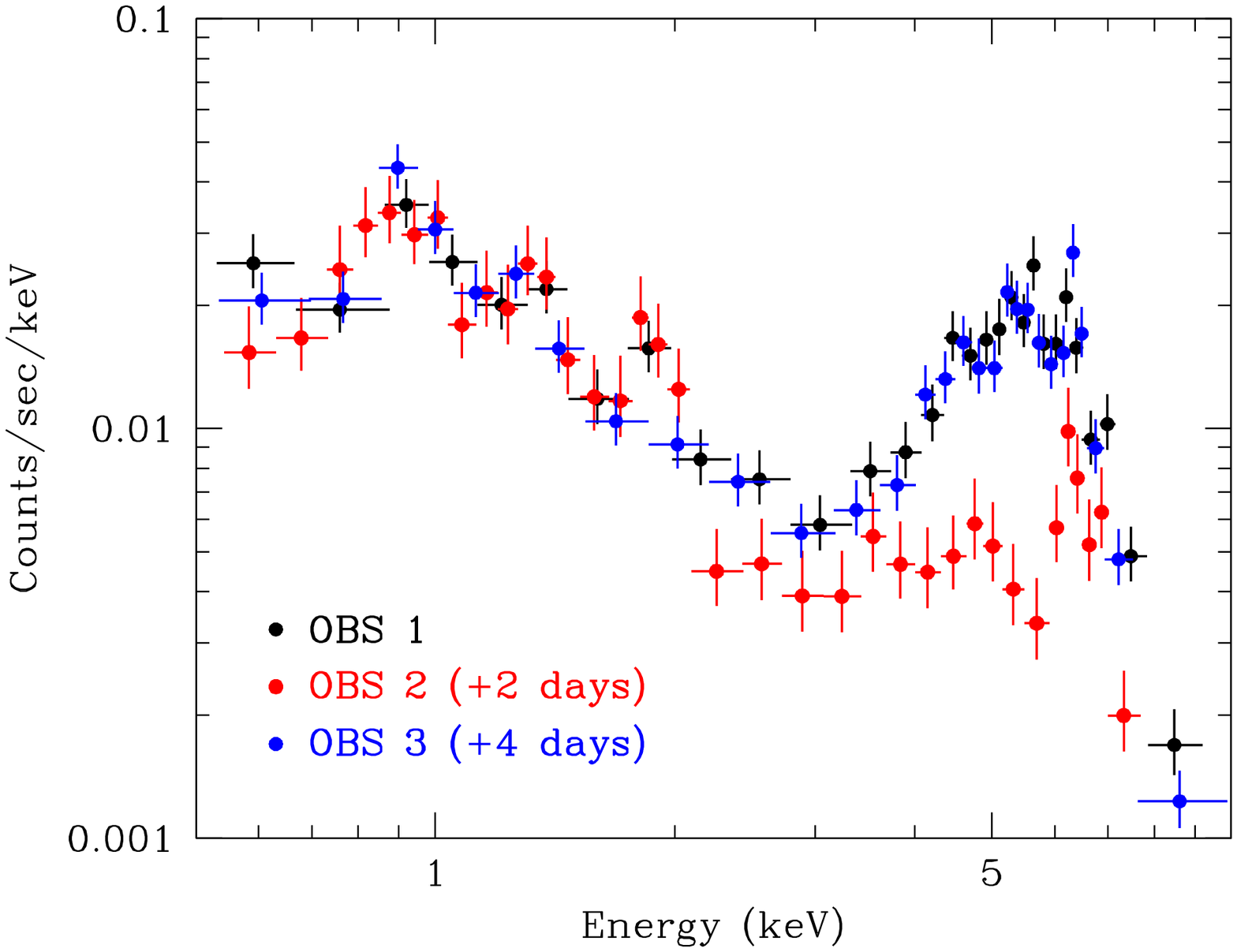}}
\hfil
\parbox{0.45\textwidth}{
\includegraphics[height=0.45\textwidth,angle=-90]{n1365eclipses2.eps}}
\hfil
}
\caption{\label{n1365eclipses}Spectra of NGC~1365, showing X-ray occultations due to BLR clouds: \textit{left}: spectra obtained from three {\em Chandra} snapshot observations performed every two days \citep{ris07}. The reflection-dominated spectrum in the second observation is due to an occultation by a Compton-thick cloud (the only one ever detected in short time scales). \textit{Right}: spectra obtained from different intervals (length of about 10~ks) of a single {\em Suzaku} observation, revealing absorption changes due to a cloud with a column density of the order of 10$^{23}$~cm$^{-2}$.}
\end{figure*}

In particular, in the case of NGC~1365, {\em Chandra}, XMM-\textit{Newton} and {\em Suzaku} observations
revealed extreme spectral changes, from Compton-thin (N$_H$ in the range 10$^{23}$~cm$^{-2}$) to
reflection-dominated (N$_H>10^{24}$~cm$^{-2}$) on time scales from a couple of days to $\sim$10~hours
(see Fig.~\ref{n1365eclipses}).  Such rapid events imply that the absorption is due to clouds with
velocity v$>$10$^3$~km~s$^{-1}$, at distances from the BH of the order of 10$^4$ gravitational radii
(assuming that they are moving with Keplerian velocity around the central black hole). The physical size
and density of the clouds are estimated to be of the order of 10$^{13}$~cm and 10$^{10}$-10$^{11}$~cm$^{-3}$, respectively. All these physical parameters are typical of BLR clouds, strongly suggesting that the X-ray absorber and the clouds responsible for broad emission lines in the optical/UV are one and the same.

These results are obtained from the analysis of the observed absorption variations in the sources mentioned above, assuming a simple scheme, where the clouds are homogeneous, with a constant column density, and are moving across the line of sight with Keplerian velocity.
However, in a few cases with particularly high signal-to-noise, the analysis of X-ray ``eclipses'' can provide further information on the geometrical and physical structure of the cloud.

\begin{figure*}
\centering
\includegraphics[width=0.98\textwidth]{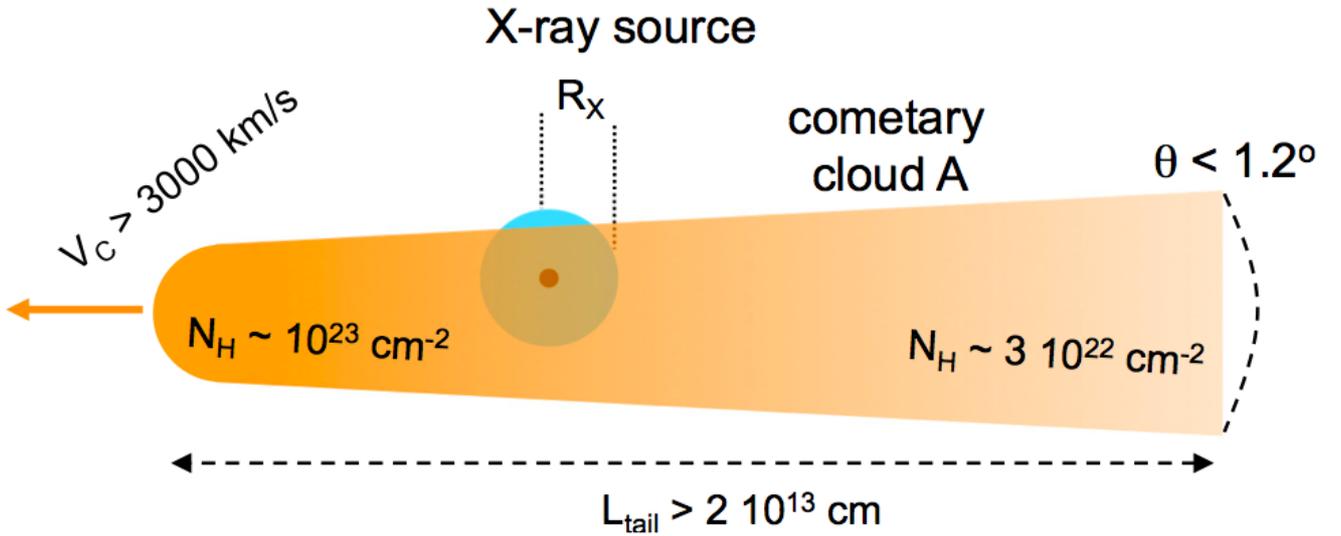}
\caption{\label{n1365comet} Structure of the absorbing cloud as obtained from a {\em Suzaku} observation of NGC~1365. The estimates are based on the hypothesis of Keplerian motion, and on a black hole mass of 2$\times$10$^6$~M$_\odot$ \citep{mai10}. The cloud size is not in the correct scale: the tail is much longer when compared with the source size, which is of the order of a few 10$^{11}$~cm. }
\end{figure*}

In the case of NGC~1365, a careful analysis \citep{mai10} of the spectral X-ray variability during two
eclipses revealed a ``cometary'' shape of the obscuring cloud, consisting of a high density head, and an
elongated, lower density tail. This structure is revealed by the time evolution of the two key
observational parameters of the cloud (Fig.~\ref{n1365comet}): its covering factor to the X-ray source
(suddenly increasing at the beginning of the occultation, then slowly increasing over a relatively long
time interval), and its column density (highest at the beginning of the occultation,
and then decreasing steadily).

Occultations have also been observed in the bright Narrow Line Seyfert 1 Mrk~766 \citep{risa11}, revealing that such events are possible (though rare) in on-average unobscured sources. While the data quality was not enough to perform a column density/covering factor deconvolution as the one described above for NGC~1365, additional information came from the detection of highly ionized iron absorption lines. These lines, due to \ion{Fe}{xxv} and \ion{Fe}{xxvi} ions, are present only in the same time intervals where the occultations due to the neutral cloud are also present. Furthermore, their energy clearly reveal that the absorbing gas is outflowing with velocities of several 10$^3$~km~s$^{-1}$. The straightforward interpretation of these measurements is that the absorption is due to an outflowing cloud, with a high-density, low ionization head, and a low-density, high ionization tail.

The above examples remain unique among the available X-ray observations of AGN, but they suggest that with future, larger area X-ray observatories, X-ray absorption variability may become a powerful, relatively standard way to directly measure the physical properties of the absorbing BLR clouds.

One of the most direct consequences of the presence of gas inside the dust sublimation radius is a decrease of the expected dust-to-gas ratio, as measured from the ratio between optical/near-IR reddening, and X-ray column density. This is a well know observational evidence in nearby Seyfert galaxies \citep[e.g.][]{mai01,mpe82}, which is therefore naturally explained in the context of X-ray absorption by BLR clouds.

The presence and structure of ``cometary'' clouds poses challenges and possible solutions for long-standing problems in modeling the BLR in AGN.
\citet{mai10} estimate that the cloud “head” loses a significant fraction of its mass through the cometary tail, which is expected to cause the total cloud destruction within a few months. If these clouds are representative of most BLR clouds (or at least the high-ionization ones), this implies that the BLR region must be continuously replenished with gas clouds, possibly from the accretion disk.
Such a ``dynamical equilibrium'' scenario would solve the problem of the long-term stability of BLR clouds, for which a convincing solution has not been found yet.

A further possible interesting aspect of the ``cometary'' structure (not yet explored in the literature)
is the reduction of the number of clouds needed to explain the smoothness of the profiles of the broad
absorption lines. High quality observations of the profile of optical broad emission lines suggest that the minimum number of clouds needed to reproduce the observed smoothness is quite high: one of the most striking examples is that of NGC~4151, with an estimated number of at least 10$^8$, based on high S/N {\em Keck} spectra \citep{arav98}. Considering that the estimated black hole mass of NGC~4151 is of the order of 10$^7$~M$_\odot$, and that the distance of BLR clouds is of the order of 10$^4$ gravitational radii, a huge cloud density is obtained, with a nearly complete occupation of the available volume.
If instead every single cloud has a cometary structure, analogous to that inferred for NGC~1365, each cloud would contribute to the observed profile with a small, but not null, width, which would greatly reduce the total number of clouds in order to reproduce the same smoothness.  

Finally, if the covering factor and the optical depth of the BLR are large enough, a significant fraction
of the iron K$\alpha$ line should be produced there. In order to test this hypothesis, NGC~7213
represents a unique opportunity. The X-ray spectrum of this source has no evidence for Compton
reflection, a unique result among bright Seyfert 1s \citep{bianchi03b,bianchi04,lob10}. The observed
neutral iron K$\alpha$ line, therefore, cannot be produced in a Compton-thick material, like the disc or
the torus. Indeed, the iron line is resolved in a \textit{Chandra} High-Energy Transmission Grating
observation, with a measured FWHM which is in perfect agreement with the value measured for the broad
component of the H$\alpha$ in a simultaneous optical observation \citep{bianchi08b}. Moreover, the observed equivalent width (EW) of the iron line is in agreement with an origin in the BLR, under reasonable assumptions on the geometrical distribution of the clouds, their covering factor, and their column density \citep{bianchi08b}. Therefore, NGC 7213
is the only Seyfert 1 galaxy whose iron K$\alpha$ line is unambiguously produced in the BLR. It is
difficult to reach similar conclusions for other objects, because of the presence of the Compton
reflection component never allows us to exclude the contribution from the more extended torus, and
because it is difficult to measure the FWHM of the iron line and compare it with the optical lines (see
also next section). In the future, high-resolution X-ray spectroscopy with micro-calorimeters and X-ray
reverberation studies will be extremely powerful in tackling this issue.

\subsection{\label{torus}Absorption form pc-scale tori}

Early evidence for a circumnuclear dusty medium on parsec, or sub-parsec scales,
as initially invoked by the Unified Model, was obtained from near-IR studies,
which revealed the presence
of very hot dust, close to the sublimation temperature, in the nuclei
of Sy1s \citep{bergmann92,alherr01,oomm99}. The dust
sublimation radius in Seyfert nuclei is on sub-parsec scales and on parsec scales at quasar luminosities.
The sub-parsec location of the hot dust emitting in the near-IR has been
confirmed by extensive reverberation observational campaigns \citep{sug06}, which also confirmed
the expected $L^{1/2}$ dependence of the sublimation radius.
The covering factor 
of the circumnuclear dusty medium, inferred from the near-IR observations,
is very
high in most of the nearby Sy1s \citep[exceeding 0.8:][]{mai07,treister08}, and generally in agreement
with the observed  type 2/type 1 ratio \citep{mr95}.

Radio \textit{VLBI} observations were the first ones to effectively image the AGN circumunclear medium on
parsec and sub-parsec scales. \citet{greenhill96} obtained \textit{VLBA} water maser images at sub-parsec
resolution of NGC~1068, revealing a rotating warped disk structure. The warping of the
maser disk orientation may indicate that the inclination of the nuclear molecular disk is
likely responsible for the large covering faction, rather than a geometrically thick torus
\citep{lawrence10}; however, one must also take into account that the maser emission does not necessarily
trace the global morphology of the circumnuclear molecular gas, but only the equatorial edge-on medium
(within $\sim 15^{\circ}$ from the line of sight) where maser amplification is highest.
\textit{VLBI} water maser observations were subsequently obtained for other AGN, finding similar
structures \citep{greenhill97,greenhill03,kond08,kond05}.
\citet{gallimore97} obtained \textit{VLBA} images of the nuclear continuum radio emission of NGC~1068. Beside
the nuclear non-thermal emission, tracing the nuclear engine, they resolved two symmetric
radio free-free blobs at a radius of $\sim$0.3 pc, which were interpreted as the inner ionized
edge of the obscuring torus.

\begin{figure*}
\centering
\includegraphics[width=0.48\textwidth]{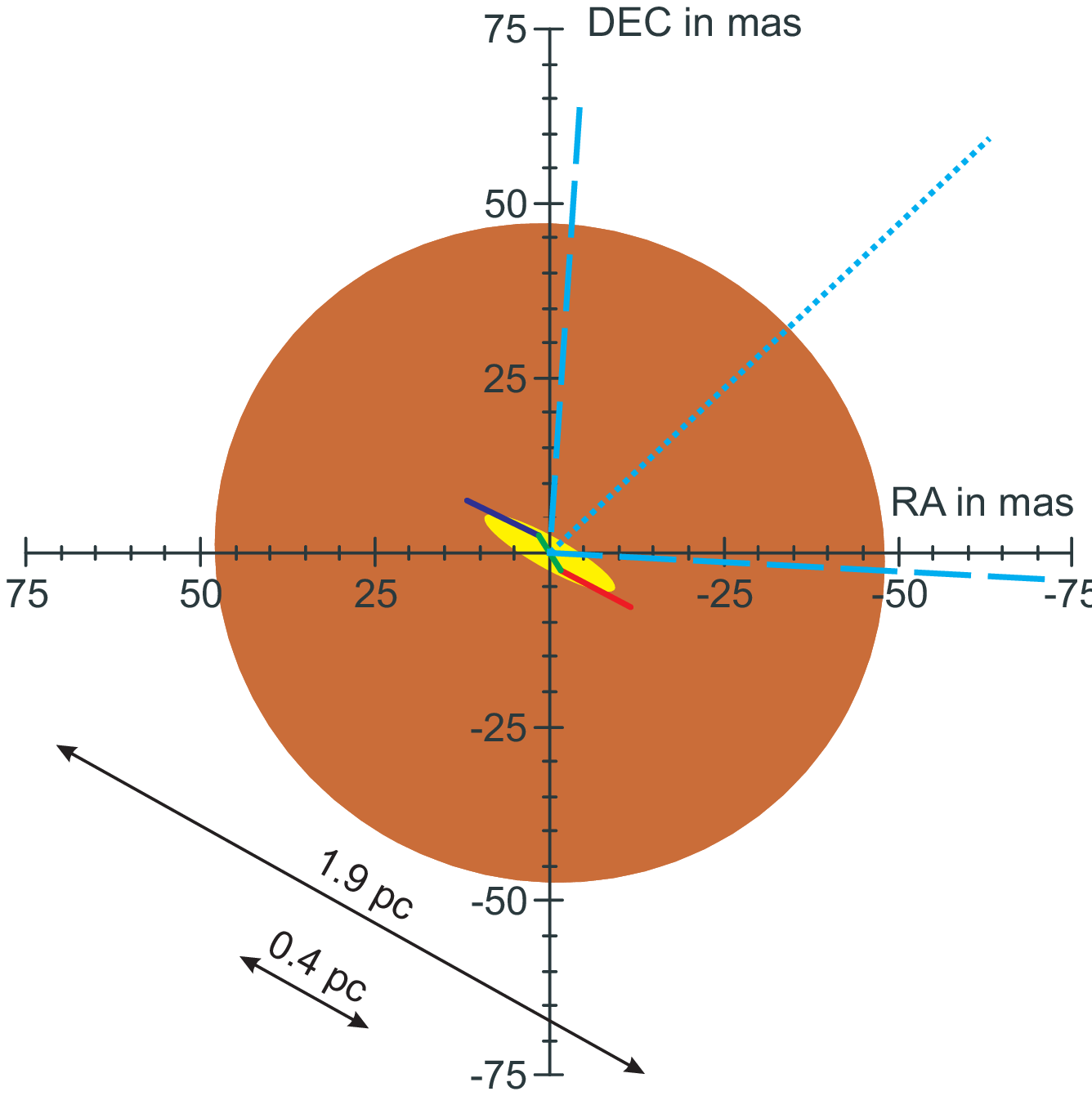}
\includegraphics[width=0.48\textwidth]{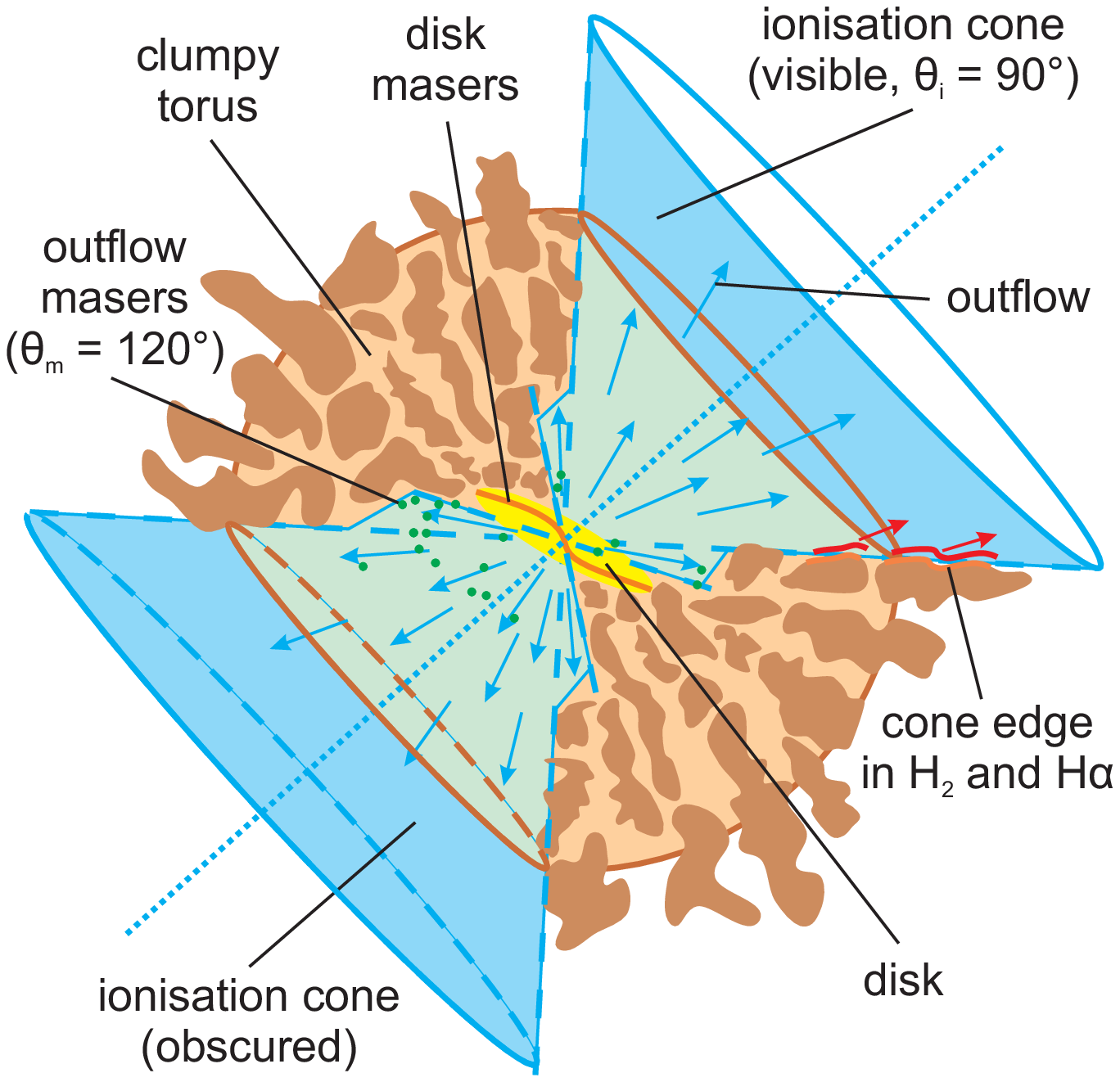}
\caption{\label{circinus_torus}The Circinus Galaxy (from \citet{tri07}). \textit{Left}: Sketch of the best-fit physical model a warm ($T\simeq330$ K) emission region (yellow) is surrounded by a cooler ($T\simeq300$ K) one (brown). The water maser emitters are over-plotted (blue and red lines), together with the ionised cone (dashed and dotted light blue line). \textit{Right}: the dusty torus as derived from the MIR interferometric observations.}
\end{figure*}

The possibility to effectively `image' the dusty component of the parsec-scale torus has become possible
in 2004, when mid-infrared interferometry allowed \citet{jaffe04} to map for the first time the dust at
parsec resolution in the Seyfert 2, NGC~1068. Their results, refined by following observations
\citep{rab09}, are consistent with a two-component dust distribution: an inner (0.5 pc of thickness),
rather elongated hot ($T>800$ K) component, and a more extended (3-4 pc), less elongated colder
($T\simeq300$ K) component. The compact component is coincident, in size and orientation, with the
nuclear water maser. Most of the absorption appears to be located outside 1 pc. A similar result was
found, with the same technique, for another Seyfert 2, Circinus: again two components, an inner and more
compact (0.4 pc), and an outer (2 pc) component \citep{tri07}. However, the temperature of the inner
component in Circinus ($T=330$ K) is significantly lower than in NGC~1068, and far from the sublimation
temperature (see Fig.~\ref{circinus_torus}). The first observation carried on a type 1 object, NGC~4151,
led to results in agreement with the previous ones on Seyfert 2s \citep{bur09}.

When such interferometric studies were performed on a sizeable sample of
objects in mid- and near-infrared, it was observed that no significant
differences are found between type 1 and 2 sources, and the size of the dusty
emitter scales with the square root of the luminosity
\citep{tri09,trisch11,kis11}. Comparisons with tori models suggest that, in
principle, it would be possible to disentangle between face-on and edge-on
distributions by comparing the compactness of the dusty structure to the AGN
luminosity, but uncertainties on the observed measures are still too large.

The presence of Compton-thick neutral material with large covering factor in the environment of AGN is
also supported by the ubiquitous presence of the iron K$\alpha$ line and the Compton reflection component
in the X-ray spectra of Seyfert galaxies \citep[e.g.][]{per02,bianchi04,bianchi09}. Although a component
broadened by strong gravity effects arising in the accretion disk is observed in at least a third of the
sources \citep[see e.g.][]{delacalleFERO,nan07}, a narrow core of the iron line is a much more common
feature. The line, typically unresolved (with upper limits of several thousands km/s for its FWHM), must
be produced far from the nucleus, either in the BLR, the torus or the NLR. Apart from single exceptional
cases (like NGC~7213: see previous section), current X-ray satellites allow us to resolve its FWHM
only in a few objects and with limited information, generally leading to inconclusive estimates on the location of the material producing the lines \citep[see e.g.][]{nan06,syw11}. Future X-ray missions taking advantage of micro-calorimeters will represent a breakthrough in this kind of analysis, allowing us to deconvolve all the components possibly present in the iron line, as routinely performed for the optical lines.

However, clues in favour of a parsec-scale distance of the material producing
the narrow iron line and the Compton reflection component already come from
the lack of variability of these features. X-ray spectra of Compton-thick
sources (i.e. obscured by a column density larger than $\simeq10^{24}$
cm$^{-2}$) are completely dominated by reflection features, and they typically
do not show any variability even on long timescales. This is particular clear
in sources where the central engine fades away for a long time interval (years),
while the reflection component (including the Fe K$\alpha$ line) remain
stable over the same time scale \citep[][]{gilli00,matt03}.
This suggests that the obscuration/reflection occurs on (at least) pc-scale, like the standard torus envisaged in the Unification Models and mapped by interferometry. In principle, the geometry and distance of the torus could be estimated by doing accurate X-ray reverberation analysis of the iron line and the Compton reflection component, in order to take into account in detail how the material reacts to the intrinsic variability of the central source. Unfortunately, such a study is extremely difficult and uncertain with current X-ray missions.

\begin{figure}
\centering
\includegraphics[width=0.45\textwidth]{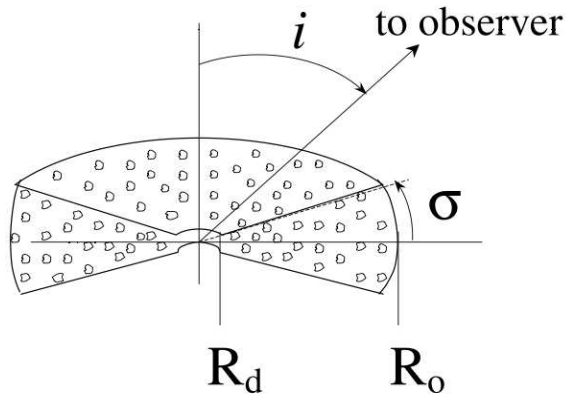}
\caption{\label{nenkova}Sketch of the dusty torus as in the model of \citet{nenkova08}, which can reproduce the observed dust temperature distributions derived from infrared observations.}
\end{figure}

While X-ray absorption variability studies have delivered exquisite information on the structure
of the absorbing medium on scales of the BLR, in particularly for what concerns the clumpiness of the
absorber, the same kind of analysis is difficult to perform on the parsec scale absorber, both
because the time scale for variation is much longer, and because it is totally overwhelmed by the
variability introduced by the clouds in the BLR. However, as mentioned in the introduction,
recent models have shown that even for the dusty pc-scale torus a clumpy structure (Fig.\ref{nenkova})
can better account for the infrared observational properties \citep{nenkova02,nenkova08,elitzur06,honig06}.
In particular, the very broad infrared Spectral Energy Distribution (SED) of AGN requires dust at multiple temperature, which is hardly
achieved by models with a compact (pc-scale) uniform torus \citep{pier92}. Large scale (100 pc) dusty
torii \citep{granato94} can reproduce the broad IR SED, but can be hardly reconciled with the
small sizes observed in mid-IR interferometric observations. A pc-scale, but clumpy torus can at the same
time match the observed mid-IR size and reproduce the broad range of dust temperatures, since within each
dense clump dust does span a wide range of temperatures. A crucial test of these models will be feasible
with \textit{ALMA}, which will allow us to image the cooler dust thermal emission of torus at sub-mm wavelengths
at sub-pc resolution.
Indeed, as discussed in \citet{maiolino08}, a clear prediction of this model is that the morphology
of the ``torus'' at far-IR/sub-mm wavelengths (tracing cold dust) should be very similar to the morphology observed
in the mid-IR (tracing warm dust). An independent observational indication of the clumpiness of the dusty
absorber is, as discussed in \citet{shi06} and in \citet{nikutta09}, the large scatter of the depth of
the 9.7$\mu$m silicate absorption feature (directly tracing
dust absorption at mid-IR wavelengths) as a function of the X-ray gaseous column density, as well
as the finding that the same feature is observed in emission in some type 2 AGN.

\subsection{\label{dustlanes}Absorption by gas in the host galaxy}

Although clear evidence in favour of absorption from the BLR and the pc-scale ``torus'' has been
presented in the previous sections, in some cases the lowest column densities are
consistent with the optical reddening associated with the medium in the host galaxy, which
is therefore an additional ingredient
that must therefore be considered in a global Unification Model \citep[see
e.g.][]{mr95,matt00b}. Early evidence of obscuration on 100 pc scales by the host galaxy gaseous
disk came from the finding that optically selected AGN samples tend to avoid edge-on systems
\citep{mr95}, a result which has been confirmed and refined with much higher statistics by using
the \textit{SDSS} survey \citep{lagos11}. Furthermore, it has been suggested that the gas in the host galaxy disk
can partially obscure also the NLR.

\begin{figure*}
\centering
\includegraphics[width=0.96\textwidth]{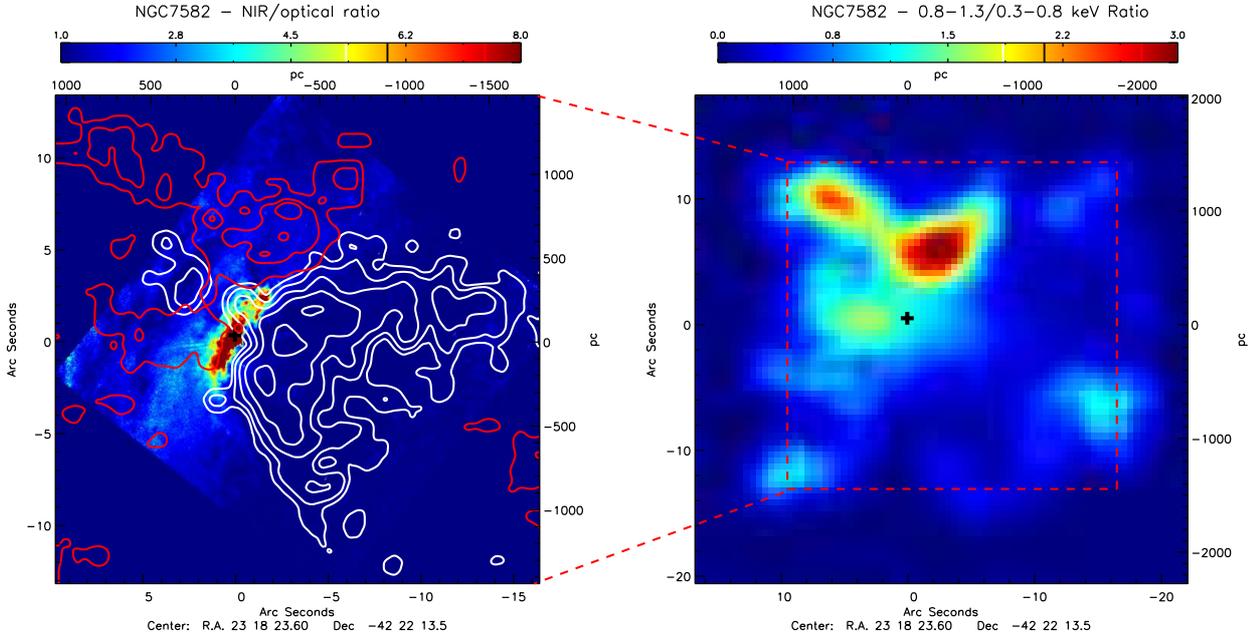}
\caption{\label{7582_dustmap}The Seyfert 2 galaxy NGC~7582 (from \citet{bianchi07b}). \textit{Left:} \textit{HST} NIR to optical ratio, mapping the amount of dust in the circumnuclear region of NGC~7582. The white contour plots refer to the \textit{Chandra} emission below 0.8 keV, while the red ones the ratio shown in the right panel. \textit{Right:} \textit{Chandra} 0.8-1.3 to 0.3-0.8 keV ratio. The red contour plots shown in left panel refer to this image, but the scale is different, as outlined by the broken-line rectangle. In both panels, the black cross indicates the position of the nucleus.}
\end{figure*}

Further direct evidence for obscuration on ``large'' scales was obtained
through high-resolution \textit{HST} images, showing that dust
lanes at distances of hundred of parsecs are very common in Seyfert galaxies
\citep{malk98}. The presence of these structures is correlated with Compton-thin X-ray
obscuration, even if not necessarily being directly responsible of the obscuration of
the nucleus \citep{gua05b}. In some cases, the effect of dust-lanes can be seen directly
as X-ray obscuration towards the soft X-ray emission from the
NLR \citep[e.g.][see Fig.~\ref{7582_dustmap}]{bianchi07b}. 

Interferometric maps of the molecular gas distribution have provided additional
evidence for large amount of dense gas on the 100 pc scale surrounding AGN, which
certainly contributes to the obscuration of the central engine along some line of sights
\citep[e.g.]{schinnerer00,boone11,krips11}.
Certainly, within this context, the advent of ALMA is going to be a breakthrough
by providing detailed maps of the molecular gas distribution in the circumnuclear region of many AGN.

As it will be discussed further below, there is evidence that the role of absorption from gas
in the host galaxy
becomes increasingly important in high-z AGN \citep{polletta08,martinez06}, likely as a consequence of
the higher gas content and higher star formation in high-z AGN hosts.

It is worth reminding that the obscuration occurring on such large scales is limited by dynamical mass
constraints. \citet{ris99} showed that Compton-thick gas must be contained on scales significantly smaller
than 100 pc, in order  not to exceed the dynamical mass in the same region and have a covering factor
large enough to account for the high number of observed Compton-thick sources. This in turns means that
the bulk of the ubiquitous Compton reflection component and narrow neutral iron K$\alpha$ line must also come from a compact region. However, in NGC~1068 hard X-ray emission (mostly associated to the reflection component) and the neutral iron line are seen extending up to $\simeq2$ kpc from the nucleus \citep{yws01}.

\section{Open issues}

\subsection{``True'' Seyfert 2s}

As discussed in Sect.~\ref{introduction}, one of the most convincing pieces of evidence in favor of the Unification Model is the detection of broad optical lines in the polarized spectra of type 2 AGN. However,
about half of the brightest Seyfert 2 galaxies appear not to have hidden broad-line
regions in their optical spectra, even when high-quality spectro-polarimetric
data are analysed \citep{tran01,tran03}.

A number of these cases may be associated
with nuclei where the mirror reflecting the broad lines either has very low scattering
efficiency (either due to
low covering factor or low column density) or is obscured \citep{heisler97}.
Evidence was also obtained that the lack of polarized broad lines is associated
with a stronger contribution/dilution from the host galaxy or from a circumnuclear starburst,
making the detection of polarized broad lines harder \citep{alexander01,gu01}.

However, a number of Sy2s without polarized broad lines may be genuine
type 2 Seyferts, in the sense that they intrinsically lack a BLR.
Indeed, observational evidence suggests that Seyfert 2s with polarized broad lines are
more easily associated with
truly obscured Seyfert 1 nuclei, while
Seyfert 2s without polarized broad lines preferentially host weak AGN, possibly
incapable of generating a classical BLR \citep{tran01,tran03}.

In a model proposed by \citet{nic00}, the creation of the BLR is connected with disk
instabilities occurring in proximity of a transition radius at which the accretion disk
changes from gas-pressure dominated to radiation-pressure dominated. Since this transition
radius  becomes smaller than the innermost stable orbit for very low accretion rates (and
therefore luminosities), very weak AGN are expected not to be surrounded by a BLR. 
More recently, the disappearance of the BLR in low-luminosity AGN has been predicted 
by \citet{ElitzurHo} and \citet{trump} within a disk-wind scenario, where the BLR is
embedded in an outflowing wind, which is no longer supported by the disk below a critical
value of the Eddington ratio.

If the BLR cannot form in weakly accreting AGN, we expect the existence of ``true'' Seyfert 2
galaxies, i.e. optically classified Type 2 objects, without any evidence of obscuration of
their nuclei. Such unabsorbed Seyfert 2 galaxies do exist, and the best examples (where the
lack of the optical broad lines and of the X-ray obscuration are unambiguously found in
simultaneous observations with high SNR) have low Eddington rates: NGC~3147 \citep[$\log
L_\mathrm{bol}/L_\mathrm{Edd}\simeq-4$:][]{bianchi08a}, Q2131−427 \citep[$\log
L_\mathrm{bol}/L_\mathrm{Edd}\simeq-2.6$:][]{panessa09}, NGC~3660 \citep[$\log
L_\mathrm{bol}/L_\mathrm{Edd}\simeq-2$: ][]{brna08,bianchiprep}. When a sizeable catalogue of
X-ray unobscured radio-quiet AGN is analysed, these sources are among those with the
lowest accretion rates \citep[e.g.][]{mar12}. It is interesting to note that, although lacking the BLR, these objects do appear to have a face-on Compton-thick torus, as type 1 sources, as evidenced by the presence of silicate emission features in their IR spectra \citep{shi10}, and of a neutral iron emission line \citep{bianchi08a}.

\begin{figure}
\centering
\includegraphics[width=0.48\textwidth]{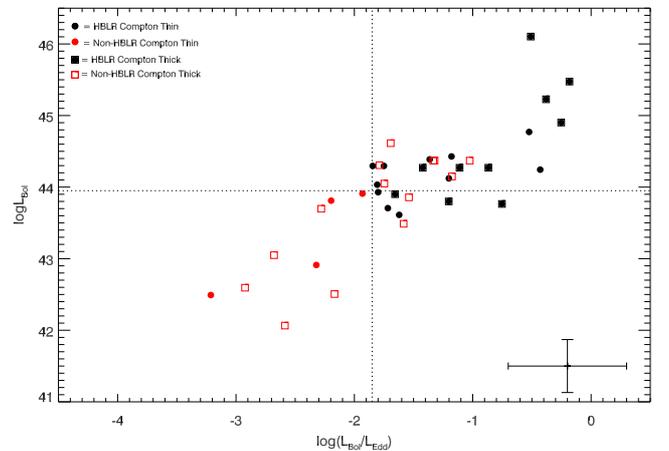}
\caption{\label{truesy2_fig1}The separation between Seyfert 2 with detection of polarized
broad lines (Hidden BLR, HBLR) and Seyfert 2s without polarized broad lines (Non-HBLR) is clear both in X-ray luminosity and Eddington rate, when only Compton-thin sources are taken into account. However, a number of Compton-thick Non-HBLR sources have high accretion rates and luminosity, suggesting that obscuration also plays an important role in the detection of the HBLR (from \citet{mar12}).}
\end{figure}

The existence of a critical threshold in luminosity and Eddington rate has been confirmed on
observational grounds \citep[e.g.][]{nic03,biangu07,shu07,wu11,mar12}. Below this threshold
(whose exact value depends on the adopted sample, and the methods to derive the bolometric
luminosities), no broad lines are detected (either in total or polarized light). On the other hand, it is clear that above the threshold the BLR still cannot be detected in many sources (see Fig.~\ref{truesy2_fig1}). If the scenario proposed by \citet{nic00} is correct, these sources should possess a BLR, so there must be something that prevents us from observing it. Indeed, all these sources are Compton-thick, so the nucleus is severely obscured by intervening absorbers. It was suggested that this could be explained within the framework of standard Unification Models, whereby more inclined sources (with respect to the line-of-sight) should intercept a larger column density of the torus, and may obscure the medium responsible for the scattering of the BLR photons \citep[e.g.][and references therein]{shu07}.

\subsection{Disk-torus alignment}

Although not explicitly required by any Unification Model, the most natural assumption on the
geometry of the circumnuclear matter in AGN is that it is co-axial with the spin of the BH. This expectation is based on an angular conservation argument: if the obscuring torus is related to the inflowing material, it is natural to expect that the torus, the accretion disk, and the black hole rotation (mostly due to the angular momentum of the accreting material) share the same axis. This hypothesis, though reasonable, may not be verified in several more complex scenarios. For example, if the BH growth is due to multiple, unrelated accretion events, the actual BH spin may not reflect the rotation axis of the accretion disk. Another possibility is that the obscuring torus is not within the gravitational sphere of influence of the BH (for example, a galactic dust lane). In this case, no obvious physical relation is expected between the torus axis and the BH spin.
If the torus-BH spin alignment hypothesis is accepted, it implies that the accretion disk is aligned with the obscuring torus. Any radio jet should be aligned to the same axis. Finally, the orientation and the opening angle of the NLR ionization cones are collimated by the inner aperture of the torus, thus being themselves co-aligned with the common disk/torus axis. These expectations are very difficult to test, due to the very small scales of the inner regions of AGN. When larger scales are imaged, at least the radio jet and the optical/X-ray NLR appear to be generally in agreement with this simple picture \citep[e.g.][]{cap96,bianchi10}.

\begin{figure}
\centering
\includegraphics[width=0.45\textwidth]{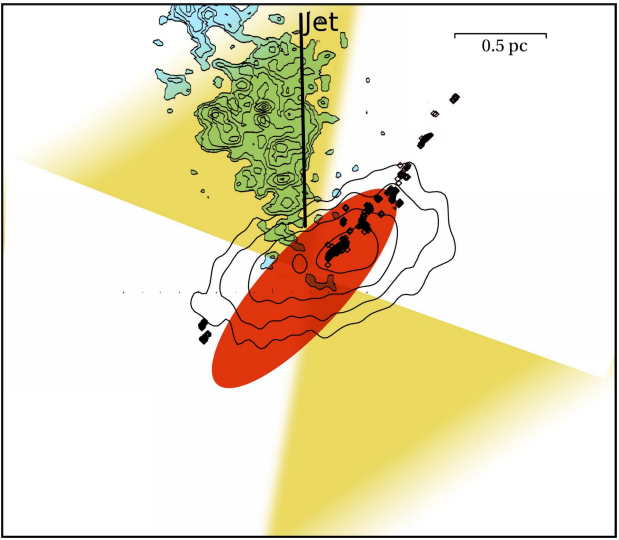}
\caption{\label{1068misalignment}Sketch of the geometry of the circumnuclear matter in NGC~1068: compact dust inferred from mid-infrared interferometry (red), ionization cones as mapped by the [\ion{O}{iii}] emission (blue, reduced in scale by a factor $\sim100$) and as inferred from spectroscopy (yellow). From \citet{rab09}.}
\end{figure}

The mid-infrared interferometric studies described in Sect.~\ref{torus} allowed us for the first time to directly image the geometry of the torus with respect to the optical cones. Surprisingly, the results obtained by \citet{rab09} strongly suggest that the two structures are misaligned in NGC~1068 (Fig.~\ref{1068misalignment}). Moreover, the direction of the radio jet is also clearly tilted with respect to both the NLR and the torus. Some of the discrepancy can be solved by taking into account the detailed kinematics of the outflow, and by assuming a clumpy torus that could prevent the ionization of all the gas present in the geometrical opening angle of the ionization cones. However, these solutions are not in agreement with the results inferred from the appearance of the cones in the infrared \citep[][and references therein]{rab09}. Similar analysis on other sources is clearly needed in order to shade some light on this issue. A promising, independent, method to test the torus/ionization cones misalignment is via X-ray polarimetry, but we will have to wait for a future X-ray mission equipped with a broad-band polarimeter \citep{gm11}.

It is more difficult to estimate the inclination angle of the accretion disk. One possible method is via the relativistic profile of iron emission lines produced in the inner regions of the disk, which is expected to be strongly dependent on the inclination angle \citep[e.g.][]{fab00}. When sizeable samples of AGN are systematically analysed, it appears that a simple relation between the inclination of the nuclear obscuring matter (as measured by
the optical type) and that of the accreting matter should be ruled out, in contrast with the naive expectations from Unified Models \citep{gua11}. However, as noted by the same authors, the inclination angles derived from the profile of a relativistically broadened line are still affected by large systematic uncertainties. Another method was recently suggested by \citet{rsm11}, who analysed the distribution of the equivalent widths of the [\ion{O}{iii}] emission line in a large sample of AGN. Their results are again not compatible with the presence of a torus co-aligned with the accretion disk, unless the torus covering factor is extremely small.

Finally, evidence of a misalignment between disk and radio jet for a few individual sources comes from the analysis of the jet-disk relation in objects where the disk inclination can be inferred from maser emission. A prominent example of this kind of studies is that of NGC~3079 \citep{kond05}.

\subsection{Luminosity and redshift dependence of the covering factor}

\begin{figure}[t]
\centering
\includegraphics[width=0.48\textwidth]{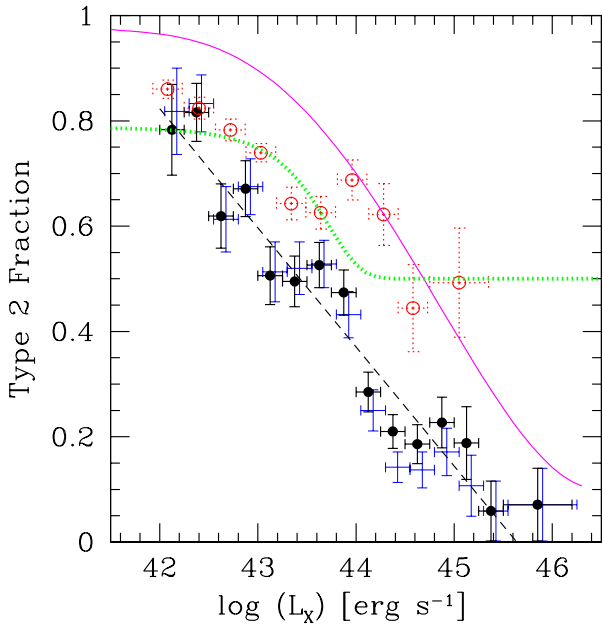}
\caption{\label{type2frac} Fraction of type 2 AGN as a function of X-ray luminosity (from
\citet{hasinger08}). The solid symbols show hard X-ray selected AGN (blue and black symbols illustrate
different sub-samples in terms of redshift completeness). The red dotted symbols show optically selected
AGN (from \citet{simpson05}). The magenta line shows the type 2 fraction inferred from the dust covering
factor obtained through near/mid-IR observations (from \citet{maiolino07}, note that these include also
Compton-thick AGN, which are not present in X-ray samples). The green dotted line shows
the expected fraction of Compton-thin type 2 AGN according to the X-ray background synthesis
models given in \citet{gilli07}).}
\end{figure}

An interesting finding of the recent years has been the evidence for a dependence
of the obscuring medium covering factor on the AGN luminosity. More specifically, the covering factor
of the obscuring medium appears to decrease significantly with luminosity.
This effect has been shown quite clearly by various hard X-ray studies 
\citep{ueda03,steffen03,lafranca05,akylas06,barger05,tozzi06} and by optical surveys \citep{simpson05},
which have measured the relative fraction of obscured and unobscured AGN as a function of the bolometric
luminosity. Clearly these works are affected by various uncertainties and caveats, primarily related
to possible incompleteness effects and biases that may prevent the identification of obscured AGN in more
distant galaxies (hence more luminous, as a consequence of the Malmquist bias). Indeed, these results
have been questioned by some authors \citep{dwelly06,treister05,wang07}. However, more recent extensive
hard X-ray surveys have further confirmed a clear trend for a decreasing fraction of obscured AGN at high
luminosity \citep{hasinger08,dellaceca08,treist09,brusa10}. Fig.~\ref{type2frac} shows the results of some of these
studies \citep{hasinger08}, where black and blue solid symbols show the fraction of obscured AGN as a
function of X-ray luminosity, while red dotted symbols illustrate the same trend from optical surveys
\citep{simpson05}.

The luminosity dependence of the covering factor has been recently questioned by \citet{lawrence10},
showing that, at least in optical and IR selected sample, the luminosity dependence is
largely a consequence of the inclusion of low excitation AGN (which may have an intrinsically
different engine) and on the definition of ``obscured" AGN (i.e. whether including mildly obscured
AGN or not). However, the luminosity dependence of the covering factor still persists in the
X-ray selected samples, regardless of the classification scheme.

An alternative method to investigate the covering factor of the obscuring medium is by means of the
dust reprocessing by the obscuring medium. More specifically, the ratio between the hot dust emission
observed in the near/mid-IR and the primary AGN bolometric emission (optical/UV/X-ray), responsible
for heating the dust, is proportional to the covering factor of the obscuring medium. This
method can only be applied to type 1 AGN, where the primary optical/UV
radiation is directly detectable.
By using \textit{ISO} and \textit{Spitzer} data to trace the dust emission, various studies have confirmed that the
covering factor of the absorbing medium decreases as a function of luminosity, as shown in
Fig.~\ref{type2frac}
\citep{wang05,maiolino07,treister08,mor11} (but see also \citep{rowan09}). These IR results obtain
a covering factor significantly higher than X-ray studies, but this is 
likely due to the population of
Compton-thick AGN that is mostly lost by X-ray surveys.

In a similar manner, the equivalent width of the (narrow) Fe K$\alpha$ line has been used as a proxy
of the covering factor of the circumnuclear material responsible for producing the Fe K$\alpha$ line
relative to the primary X-ray continuum. The EW of the Fe K$\alpha$ has also been found to anti-correlate with luminosity
(``X-ray Baldwin effect'' or ``Iwasawa-Taniguchi effect''), a trend which was generally interpreted
in terms of decreasing covering factor of the circumnuclear absorbing medium as a function of luminosity
\citep{iwasawa93,page04,jiang06,guainazzi06,bianchi07}.

Therefore, although partly questioned, several independent observational studies strongly favor a luminosity dependence of the covering factor of the circumnuclear absorbing medium.
The origin of the anti-correlation between luminosity and covering factor is unclear. The ``receding
torus'' scenario \citep{lawrence91} is often invoked to explain this trend: higher luminosities imply a larger
dust sublimation radius  and, if the torus has a constant height as a function of radius,
this results into a smaller covering factor of the dusty medium. However, this scenario cannot explain
the results on the decreasing covering factor inferred from X-ray studies, which do not trace the
dusty component of the absorber.

In the model proposed by \citet{lamastra06}, the covering factor--luminosity anti-correlation
naturally arises if the X-ray obscuration in Compton-thin sources is due to interstellar gas, distributed
in a rotationally supported disk with an extension of a few hundred pc. The covering factor of this disc
diminishes as the gravitational pull from the central SMBH and the bulge increases with the BH mass (and,
therefore, the luminosity), producing the observed anti-correlation. However, this model can explain the
anti-correlation only for Compton-thin sources.

Another possible scenario is that the lower covering factor in luminous AGN (quasars) is simply
a consequence of the stronger AGN radiation pressure impinging onto the circumnuclear medium and expelling
larger fractions of material. In support of this scenario growing evidence for massive outflows
in luminous AGN has been reported in the recent years \citep{fischer10,feruglio10,rupke11,sturm11,canodiaz11}.

Some studies have also claimed an evolution of the covering factor with redshift. More specifically, by
using large samples of X-ray selected objects,
\citet{lafranca05} and \citet{hasinger08} have found evidence for a strong increase of the fraction
of obscured AGN from z=0 to z=2. This result is however more debated, due to the
larger uncertainties in disentangling luminosity and redshift effects in flux limited samples and
also more critically subject to selection effects in high-z objects. Indeed,
other authors did not find significant redshift dependence \citep{ueda03,gilli07},
while \citet{ballantyne06} and \citet{treister06} found a shallow redshift dependence of the fraction of
obscured AGN. Here we only mention that an increase of the fraction of obscured AGN at high redshift
is naturally expected by the larger gas content found in high-z galaxies \citep{tacconi10,daddi10} and
associated with the enhanced star formation rates in high-z galaxies, also observed in high-z AGN
host galaxies \citep{shao10,shi09}.

A possible additional complication to the simple version of the Unified Model
is that the covering factor of the obscuring medium
may not only depends on the AGN luminosity, and
possibly on redshift, but may also be intrinsically different between type 2 and type 1 AGN.
\citet{ricci11} have shown that the X-ray cold reflection ``hump'' in type 2 AGN
is stronger than in type 1 AGN, suggesting a larger covering factor of the
circumnuclear medium in the former class. This result is however at odds with those obtained
by mid-IR interferometry studies, as discussed above.
More data are certainly required to tackle the discrepancy.

\subsection{Unusual geometries}

While the general picture discussed in this review can account for most of the observational
properties observed in AGN, some sub-samples require different geometries. This is the case
for a number of nuclei that are unambiguously hosting a relatively powerful AGN, based on their
X-ray or mid-IR properties, but which do not show any evidence for classical NLR tracers in their
optical spectra \citep{marconi00,dellaceca02,maiolino03,ballo04,caccianiga07,
imanishi10,imanishi07,imanishi09,risaliti10,nardini09,
nardini08}. A possibility is that in these objects the NLR, although extended,
is heavily obscured by absorbing medium distributed on large scales in the host galaxy. Indeed,
in a number of objects \textit{Spitzer} spectroscopic observations have revealed mid-IR high excitation
lines (e.g. [NeV]14.3$\mu$m) typical of the NLR \citep{evans08,goulding09,perez11}.

However, a number of active nuclei do not shows NLR-like lines even in the mid-IR \citep{armus07}.
In these cases a likely scenario is that the nuclear engine is obscured in all directions ($4\pi$
obscuration) on small ($\sim$pc) scales, so that UV photons cannot escape to produce a NLR.
Further observational indications in favour of $4\pi$ obscuration is the 
very small amount of reflection component at energies below 10~keV in some Compton thick AGN \citep[e.g.][]{ueda07,nta09}.
The most likely explanation in this case is that the Compton thick absorbing medium totally cover
the nuclear sources, including the Compton reflecting medium.

If totally ``buried'' AGN may be one extreme of the population, on the other extreme
there is growing evidence for a population of (type 1) AGN missing the circumnuclear absorber.
\citet{jiang10} discovered two quasars at z$\sim$6 showing no indication of hot dust emission
typically observed in AGN. They interpreted this result as evidence for young quasars that
had no time yet to form dust at such early epochs. However, evidence for AGN with little or no
circumnuclear hot dust emission has been found also at lower redshifts \citep{hao10,hao11,mor11},
in evolved quasars, where timescale for dust formation is likely not an issue. The nature of
this class of objects is still unclear. Possible scenarios invoke dust destruction (dynamically
or by radiation) or the AGN is not centered onto the supermassive black hole, as it may happen
in the case of BH recoils in merging events, which may result in the off-nuclear AGN to be still
surrounded by an accretion disk and a BLR, but not from pc-scale dusty medium \citep{hao11}.

\section{Summary and conclusions}

The Unified Model for AGN has been tested in many different ways in the past few years, through a 
large set of new imaging, spectral, and timing observations. Overall, the fundamental aspect of the model, i.e. that non-spherically symmetric absorption plays a major role in explaining the differences in the observed features among AGN, has been confirmed, and even reinforced by the most recent observations.

However, some more complexity has been added. In particular, the existence of the standard ``torus'', in the generic sense of an axisymmetric, rather than spherically symmetric, circumnuclear absorber has been confirmed, but its physical and geometrical structure has been proven to be far from homogeneous among the AGN family. There is now strong evidence of at least three absorption components on very different scales:\\
- on scales of hundreds of parsecs, or even larger (e.g. galactic dust lanes), circumnuclear tori have been imaged, with different techniques, and are clearly responsible of the ``type~2" (in optical/UV) or ```absorbed'' (in X-rays) classification of a significant fraction of AGN.\\
- on the parsec scale, and down to the dust sublimation radius, the ``standard'' torus, as initially postulated in the earliest works on AGN unification, has been now directly imaged in a few sources with interferometric techniques, and its presence is suggested by X-ray reflection properties, and by dust reverberation mapping in the near-IR.\\
- on the 0.01~pc scale, the presence of dust-free gas along the line of sight has been demonstrated through X-ray absorption variability in several AGN, thus suggesting that part of the observed X-ray absorption is due to Broad Line Region clouds.

From the latter study, and from models of dust re-emission in the infrared, it is also clear that the absorber has a clumpy, rather than homogeneous, structure.

Several ``open issues'' have been also briefly discussed in this review, showing that not all AGN fit in a simple unification scheme, and that even within the general boundaries of the model, more work and new observations are needed to fully understand the dependence of all the observed properties on the physical parameters of the central source and of its environment.

\subsection*{Acknowledgments}
We would like to thank the anonymous referees for useful suggestions, and R. Antonucci, M. Elvis, G. Matt, K. Tristram for feedback on the first version of the manuscript.

\bibliographystyle{apsrev}
\footnotesize
\setlength{\bibsep}{3pt}
\bibliography{sbs}

\end{document}